\documentclass[prl,twocolumn]{revtex4}

\usepackage{amsfonts}
\usepackage{amsmath}
\usepackage{amssymb}
\usepackage{graphicx}

\begin{document}

\title{Evidence for two fluids in the ground state of cuprates}
\author{Amit Keren}
\author{ Amit Kanigel}
\altaffiliation[Current Address: ]{Department of Physics, University of Illinois at Chicago, IL 60607}
\author{Galina Bazalitsky}
\affiliation{Technion - Israel Institue of Technology.}

\begin{abstract}
We report charge density measurements, using NMR, in the superconducting
compound (Ca$_{x}$La$_{1-x}$)(Ba$_{1.75-x}$La$_{0.25+x}$)Cu$_{3}$O$_{y}$,
which has two independent variables $x$ (family) and $y$ (oxygen). For
underdoped samples we find the rate at which holes are introduced into the
plane upon oxygenation to be family-independent. In contrast, \emph{not} all
carriers contribute to either antiferromagnetic or superconducting order
parameters. This result is consistent with a two fluid phenomenology or
intrinsic mesoscopic inhomogeneities in the bulk. We also discuss the impact
of weak-chemical-disorder on $T_{c}$.
\end{abstract}

\date{\today }
\maketitle

Proper counting of holes in the cuprates is essential for understanding
their properties. Since they have a small ratio of coherence length to mean
free path, they are considered clean superconductors where the carrier
density $n$ is the same as\ the superconducting carrier density $n_{s}$ in
the zero temperature limit \cite{Tallon}. In contrast, modern experiments in
the superconducting part of the phase diagram consistently find strong
inhomogeneities in these materials \cite{InhomExp,KerenSSC03}. Moreover,
several theories of cuprate superconductivity are based on two fluids
comprised of: hole-poor and hole-rich regions \cite{EmeryPhysicaC}, bosons
and fermions \cite{Boson}, hot and cooled electrons \cite{Pines}, etc.. This
contradiction leads to the question: do all holes participate in the
superconducting order parameter? Addressing this question requires a
simultaneous measurement of $n$ and $n_{s}$. The in-plane $^{63}$Cu(2)
nuclear quadrupole resonance (NQR) parameter $\nu _{Q}$ is a direct measure
of $n$. $n_{s}$ can be extracted from the penetration depth. In this work we
compare the two numbers for different superconducting families. We find that 
$n_{s}$ is \emph{not} a universal fraction of $n$, and provide a simple
relation between $n_{s}$ and $y$, from parent to overdoped samples. The data
support two fluids-based theories for the entire phase diagram, but
constrain the possible division between the fluids.

\begin{figure}
\begin{center}
\includegraphics[width=7.5cm]{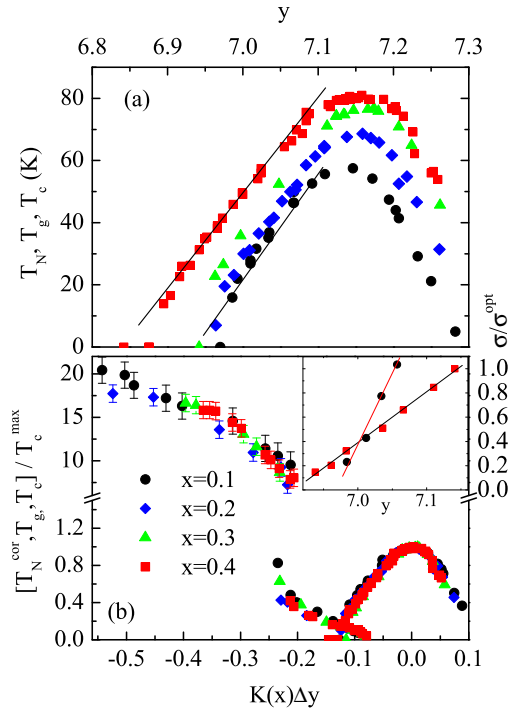}
\end{center}
\caption{(a) The superconducting part of the (Ca$_{x}$La$_{1-x}$)(Ba$%
_{1.75-x}$La$_{0.25+x}$)Cu$_{3}$O$_{y}$ phase diagram \protect\cite{clblco}.
(b) Normalized critical temperatures plotted as a function of $K(x)\Delta y$%
, including the glass $T_{g}$ temperature, and corrected N\'{e}el
temperature $T_{N}^{cor}$ (see text) \protect\cite{OferToBe}. Inset, the
normalized muon spin relaxation rates taken from Ref.~\protect\cite%
{KerenSSC03} as a function of oxygen level $y$ for the $x=0.1$ and $0.4$
CLBLCO samples.}
\label{TcandUni}
\end{figure}

Our study is done on the (Ca$_{x}$La$_{1-x}$)(Ba$_{1.75-x}$La$_{0.25+x}$)Cu$%
_{3}$O$_{y}$ (CLBLCO) compound where each value of $x$ is a superconducting
family with its own maximum $T_{c}$ ($T_{c}^{max}$) ranging between $58$ and 
$80$~K, as shown in Fig.~\ref{TcandUni}(a) \cite{OferToBe}. By choosing a
sample with a particular $x$ and oxygen level $y$ one can control $n$ and $%
n_{s}$ independently. This allows determination of the rates at which
oxygenation produces carriers and carriers turn superconducting. We evaluate
the average $\nu _{Q}$ form $^{63}$Cu(2) nuclear magnetic resonance (NMR)
measurements. In passing, we estimate the chemical disorder from the width, $%
\Delta \nu _{Q}$, of the $\nu _{Q}$ distribution. $n_{s}$ is obtained from
previous muon spin rotation ($\mu $SR) measurements \cite%
{KerenSSC03,OferToBe}.

The ability of NMR to determine carrier density is based on the fact that $%
^{63}$Cu, with its spin $3/2$ nuclei, is directly coupled to charge degrees
of freedom via the electric field gradient (EFG), and $\nu _{Q}$ is a
measure of this coupling. $\nu _{Q}$, in turn, depends linearly on the hole
density \cite{Asayama} according to 
\begin{equation}
\nu _{Q}=An+{}\nu _{Q}^{0}\text{,}  \label{nuqvsn}
\end{equation}%
where $A$ and $\nu _{Q}^{0}$\ are doping-independent, but, in principle,
could be family-dependent. This linear dependence was demonstrated for
various compounds such as Y123 \cite{Yasuoka}, La$_{2-x}$Sr$_{x}$CuO$_{4}$
(La$214$) \cite{Zheng}, and HaBa$_{2}$CuO$_{4+\delta }$ (Ha124) \cite%
{Gippius}. Therefore, Eq. \ref{nuqvsn}\ and the ability of NMR to detect the
in-plane copper [Cu(2)] $\nu _{Q}$ selectively will allow us to determine
the evolution of the in-plane carrier concentration and the width of its
distribution.

\begin{figure}
\begin{center}
\includegraphics[width=9cm]{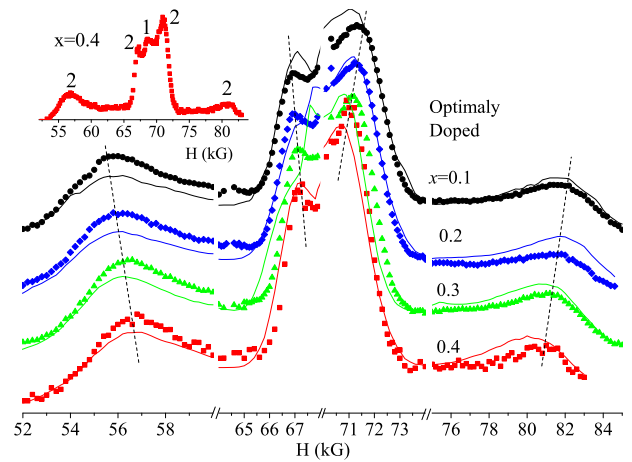}
\end{center}
\caption{NMR spectra of $^{63}$Cu at $T=100$~K in optimally doped CLBLCO
samples with varying $x$. The inset shows the full spectrum of the $x=0.4$
compound including contributions from Cu(1) and Cu(2). The main figure zooms
in on the Cu(2) contribution (note the three axis breakers). The position of
the Cu(2) peaks are shown by dotted lines.}
\label{Optimalfit}
\end{figure}

The measurements were done on powder samples fully enriched with $^{63}$Cu.
Their preparation is described in Ref.~\cite{clblco}. The oxygen content was
measured by double iodometric titration. The accuracy of this method in the
enriched CLBLCO is about 0.01. We measured between five and seven different
samples for each $x$ in the normal state at 100~K. The most overdoped sample
is a non superconducting $x=0.1$ compound. The NMR measurements were done by
sweeping the field in a constant applied frequency $f_{\text{app}}=77.95$%
~MHz, using a $\pi /2$ - $\pi $ echo sequence. The echo signal was averaged $%
100,000$ times and its area evaluated as a function of field. The full
spectrum of the optimally doped $x=0.4$ sample ($y=7.156$) is shown in the
inset of Fig.~\ref{Optimalfit}.

The Cu spin Hamiltonian can be written as \cite{Slichter}: 
\begin{equation}
\mathcal{H}/h=-\nu _{l}\mathbf{I\cdot (1+K)\cdot \hat{H}}+\frac{\nu _{Q}}{6}%
[3\mathbf{I}_{z}^{2}-\mathbf{I}^{2}+\eta (\mathbf{I}_{x}^{2}-\mathbf{I}%
_{y}^{2})],  \label{H}
\end{equation}%
where $\nu _{l}=(^{63}\gamma /2\pi )H$, $\hat{\mathbf{H}}$ is a unit vector
in the direction of the field, $\mathbf{K}$ is the shift tensor, and $\eta $
is the asymmetry parameter of the EFG. In the absence of magnetic field,
there is only one transition frequency given by $f=\nu _{Q}\sqrt{1+\eta /3}$%
, so $\nu _{Q}$ cannot be separated from $\eta $, and the use of the
magnetic field is essential. This field, applied in the direction $\theta $
and $\phi $ with respect to the principal axis of the EFG, lifts this
degeneracy, and three transition frequencies $\nu _{m}(H,\theta ,\varphi )$
are expected: a center line which corresponds to the $1/2\rightarrow -1/2$
transition ($m=0$), and two satellites which correspond to the $%
3/2\rightarrow 1/2$ ($m=1$) and $-1/2\rightarrow -3/2$ ($m=-1$) transitions.
Expressions for $\nu _{m}(H,\theta ,\varphi )$ up to second order
perturbation theory in $\nu _{Q}$ for completely asymmetric EFG and shift
tensors are given in Ref.~\cite{Taylor}. In a powder spectrum, where $\theta 
$ and $\varphi $ are integrated out, each one of the $m=-1$ and $m=1$
transitions contributes one peak, and the $m=0$ transition generates two
peaks provided that $\eta <1$ \cite{Taylor,KerenPRB98}. In principle, the
bigger $\nu _{Q}$ is, the further the peaks are away from each other.
Similarly, the wider the peaks are, the broader the distribution of $\nu
_{Q} $. In CLBLCO as in YBCO, the middle peak is from the Cu(1) which has $%
\eta \sim 1$. This peak is labeled $1$ in the inset of Fig.~\ref{Optimalfit}%
. The other four peaks are associated with the plane Cu(2) and are labeled $%
2 $.

A zoom on the main features of the Cu(2) signal of all four optimally doped
samples is depicted in Fig.~\ref{Optimalfit} (note the three axis breakers).
The evolution of the main peaks as $x$ increases is highlighted by the
dotted lines. It is clear that as $x$ decreases the peaks move away from
each other. This means that $\nu _{Q}$ at optimal doping is a decreasing
function of $x$. A more interesting observation is the fact that there is no
change in the width of the peaks, at least not one that can easily be
spotted by the naked eye. This means that the distribution of $\nu _{Q}$ is $%
x$-independent and that there is no difference in the weak-chemical-disorder
(WCD) between the optimally doped samples of the different families. By WCD
we mean disorder that is not strong enough to \textquotedblleft wipe
out\textquotedblright\ the contribution of the Cu(2) in its vicinity from
the spectrum. Thus, WCD is not relevant to the variation of $T_{c}^{max}$
between the different families. As we shall see, this conclusion is
supported by more rigorous analysis.

Although we limit our conclusion to WCD, it is important to place a few
limitations on the possible existence of strong chemical disorder. The
difference in ionic size between La and Ba in the Ba layer creates local
lattice distortions. These impact the apical oxygen O(4) position, which
contributes to $\nu _{q}$. However a simple calculation based on Eqs. 4 and
5 in Ref.~\cite{Chmaissem} shown that the probability of finding an O(4) out
of its ideal place increases with $x$. Therefore, if the O(4) displacement
were important for chemical disorder we would have expected the $x=0.4$
family to have a lower $T_{c}$ than the $x=0.1$, in contrast to observation.
In addition, magic angle spinning (MAS) Ca NMR found only one line. This
ensures the existence of only one Ca site. Also, since Ca and Ba have the
same valance, and the total amount of La is fixed, doping is done only by
the chain layer oxygen, which is relatively remote from the CuO$_{2}$ layer.
Therefore, chemical disorder is expected to be minimal. Overall, we cannot
find evidence for the existence or importance of strong chemical disorder.

\begin{figure}
\begin{center}
\includegraphics[width=9cm]{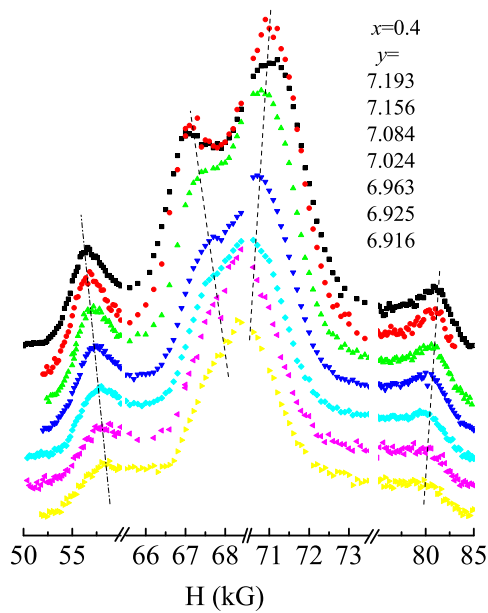}
\end{center}
\caption{NMR spectra of $^{63}$Cu(2) at $T=100$~K in CLBLCO samples with $%
x=0.4$ and varying $y.$ The positions of the peaks are shown by dotted lines.
}
\label{x04lines}
\end{figure}

The evolution of the main peaks for $x=0.4$ as a function of $y$ is shown in
Fig.~\ref{x04lines}. Here, as $y$ increases the peaks moves away from each
other, \textit{i.e.}, $\nu _{Q}$ increases as a function of doping as
expected from Eq.~\ref{nuqvsn}. Similar data for the $x=0.1$ family can be
found in Ref.~\cite{KanigelToBe}.

In order to analyze these data more precisely we must fit them to a
field-swept powder spectrum $P(H)$ which is given by%
\begin{equation}
P_{m}(H)\propto \sum_{m}\int M^{2}\delta \lbrack f_{\text{app}}-\nu
_{m}(H,\theta ,\varphi )]d\Omega  \label{trans_cond}
\end{equation}%
where $M$ is a matrix element. Usually one can extract all the averaged
Hamiltonian parameters from the position of the peaks in the spectra \cite%
{KerenPRB98}. However, we are interested in both the parameters and their
distribution. Therefore, we must account for the entire line shape. For this
purpose we simulate the line by using a grid in the $(\theta ,\phi )$ space,
calculate the frequency $\nu _{m}(H,\theta ,\varphi )$ for every $m$, field,
and point on the grid, and add 1 to a histogram of $H$ when one of the
frequencies $\nu _{m}$ equals $f_{\text{app}}$. The matrix elements are
taken as unity. This numerical simulation approximates $P(H)$ in Eq.~\ref%
{trans_cond}. To account for the peak widths we assume that the main
contribution to this width, of the order of a few MHz, is from a
distribution in $\nu _{Q}$, since the quadrupole interaction is the only
interaction of such magnitude in the system. Consequently, we added to the
numerical evaluation of Eq.~\ref{trans_cond} a loop over $200$ values of $%
\nu _{Q}$ drawn from a normal distribution with a width $\Delta \nu _{Q}$.
We also added the contribution of the chain site, with $\eta \sim 1$. We
searched for the best fit to the data by $\chi ^{2}$ minimization using a
simplex code \cite{NumericalRec}. The result for the optimal doped samples
is shown as the solid line in Fig.~\ref{Optimalfit}.

The fit is not perfect, mostly because the spectrum is not symmetric. Such
an asymmetric spectrum is a result of correlations between different \textit{%
a priory} random parameters in the Hamiltonian \cite{HasseJS00}. For
example, if the shift tensor $K$ is also non uniform in the sample but its
values are correlated with the values of $\nu _{Q}$, the spectrum could be
non symmetric. Attempts to take this kind of effect into account failed due
to the enormous increase in computing time. We continue the discussion based
on the best fit we could practically achieve. For a discussion on the error
bars evaluation see Ref.~\cite{KanigelToBe}. We could not determine, $K_{z}$%
, $K_{x}$ and $K_{y}$ very accurately, and found very small $\eta $ for all
the samples, ranging from 0 to 0.1, in agreement with various estimates for
Y123 \cite{eta}.

\begin{figure}[tbp]
\begin{center}
\includegraphics[width=9cm]{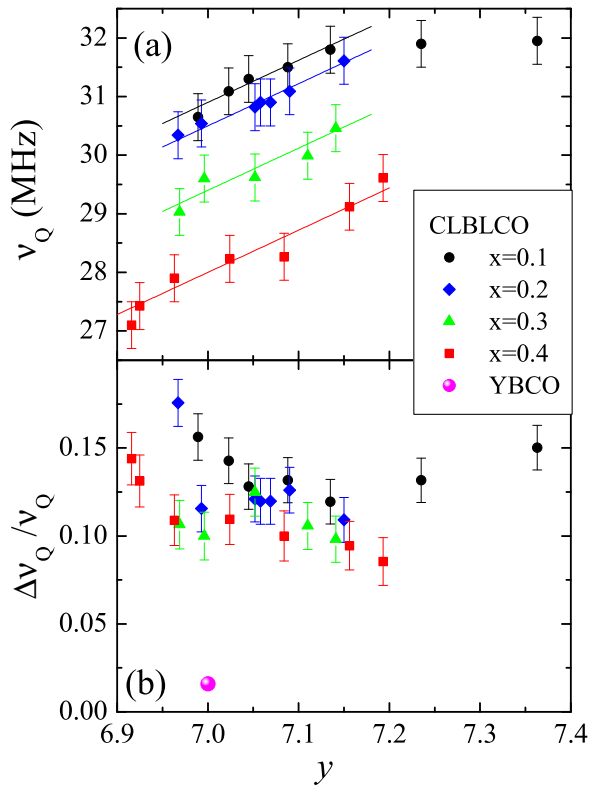}
\end{center}

\caption{(a) Nuclear quadrupole resonance frequncy of $^{63}$Cu(2) in all
the CLBLCO samples extracted from the NMR spetra as described in the text.
(b) The width of the nuclear quadrupole resonance frequency distribution of $%
^{63}$Cu(2) in all the CLBLCO samples extracted from the NMR spectra as
described in the text.}
\label{NuqDnuq}
\end{figure}

The results for fitted $\nu _{Q}$ and $\Delta \nu _{Q}$ are shown in Fig.~%
\ref{NuqDnuq} (a) and (b), respectively. From Fig.~\ref{NuqDnuq}(a) it is
clear that $\nu _{Q}$ grows linearly with doping in the underdoped side of
the phase diagram, in agreement with other compounds \cite%
{Yasuoka,Zheng,Gippius}, and Eq.~\ref{nuqvsn}. It is also clear that $\nu
_{Q}^{0}$ in this equation is $x$-dependent, but this could be attributed to
NQR base line properties. In contrast, the behavior in the overdoped side of
the $x=0.1$ is surprising since $\nu _{Q}$ saturates. As pointed out before
in Ref.~\cite{KanigelToBe}, the added holes on the overdoped side no longer
go into the planes.

The most interesting finding is that within the experimental error, the
slope of $\nu _{Q}(x,y)$ in the underdoped side is $x$-independent, as
demonstrated by the parallel solid lines, and $\partial \nu
_{Q}(x,y)/\partial y=7.2(1.4)$ MHz/Oxygen. This means that the rate at which
holes are introduced into the CuO$_{2}$ planes, $\partial n/\partial y$, is
a constant independent of $x$ or $y$ in the underdoped region. Using further
the ubiquitous assumption that the optimal hole density, $n^{opt}$, at
optimal oxygenation, $y^{opt}$, is universal, we conclude that $n(x,y)$ is a
function only of $\Delta y=y-y^{opt}$. This result was predicted previously 
\cite{Chmaissem} based on bond valance calculations.

An implication of this conclusion is that the Presland \textit{et al.}
formula~\cite{PreslanPhysicaC91} $T_{c}/T_{c}^{\max }=1-82.6(n-0.16)^{2}$
could not be correct for CLBLCO. For example, the $x=0.1$ and $x=0.4$
samples with $T_{c}=0$ have different $\Delta y$, and according to our
experiment different $n$. However, the Presland formula predicts the same $n$%
.

If the conversion from oxygen to holes is universal, but different families
begin to superconduct at different $y$ values (see Fig.~\ref{TcandUni}), it
means that the rate at which holes turn into superconducting carriers is
family-dependent. To demonstrate this concept further we focus on the region
of the CLBLCO phase diagram where $T_{c}$ grows linearly with doping, as
emphasized by the solid lines in Fig.~\ref{TcandUni}(a). This behavior is
unique to CLBLCO, which does not show a $1/8$ dip or plateau in $T_{c}$. We
determine $n_{s}$ from $\mu $SR measurement at $T\rightarrow 0$. The $\mu $%
SR relaxation rate $\sigma $ is a measure of the density of superconducting
carriers, and $\sigma /\sigma ^{opt}=n_{s}/n_{s}^{opt}$. In the inset of
Fig.~\ref{TcandUni}(b)\ we depict $\sigma /\sigma ^{opt}$, taken from Ref.~%
\cite{KerenSSC03}, as a function of $y$ for two extreme families with $x=0.1$
and $0.4$. In this particular region $\partial n_{s}/\partial y$ is a
constant, which is family-dependent and denoted hereafter by $K(x)$. Using
the universality of $\partial n/\partial y$ we find that $\partial
n_{s}/\partial n=K(x)(\partial n/\partial y)^{-1}$, which varies between
families. This, again, rules out the possibility that $n_{s}=n$ in all
CLBLCO samples.

In light of previous work a stronger conclusion could in fact be extended to
the entire phase diagram of CLBLCO. In Fig.~\ref{TcandUni}(b), taken from
Ref.~\cite{OferToBe}, we depict $T_{N}^{cor},T_{g}$ and $T_{c}$, normalized
by $T_{c}^{\max }$ of each family. $T_{N}^{cor}$ is the N\'{e}el temperature
after the contribution from anisotropies have been divided out. The scaled
variable $K(x)\Delta y$, with $K=0.77,$ $0.67$, $0.54$, $0.47$, for the $%
x=0.1$ to $0.4$ respectively, collapsed the entire phase diagram into a
single function. However, no interpretation was given for $K(x)$. The
present experiment suggests that $\partial n_{s}/\partial y=K(x)$ over the
entire phase diagram, and that $n_{s}$ should be considered as carriers
participating in both superconducting and antiferromagnetic order parameters.

As for the line width, examination of Fig.~\ref{NuqDnuq}(b) reassures us of
our previous intuition that $\Delta \nu _{Q}$ is not changing between the
different families, especially for the optimally doped samples. This becomes
obvious when considering $\Delta \nu _{Q}$ of YBCO$_{7}$ at $T=100$~K, which
also has $\nu _{Q}=31$ MHz, but $\Delta \nu _{Q}=0.5$~MHz \cite{OferPRB06}.
The changes in $\Delta \nu _{Q}$ between $x=0.4$ and $x=0.1$ are minute
compared to the changes between $x=0.1$ and YBCO$_{7}$. Yet $x=0.4$ has a $%
T_{c}^{max}$ very similar to YBCO$_{7}$. Thus, if YBCO is considered
disorder free, the difference in WCD between $x=0.4$ and $x=0.1$ either does
not exist or is not relevant. The same conclusion was reached by MAS Ca NMR 
\cite{MarchandThesis}.

In summary, we find that the rate at which holes are doped into the planes
when oxygen is added, $\partial n/\partial y$, is identical in all families
in the superconducting underdoped region. In part of this region, the rate
at which holes contribute to the condensate $\partial n_{s}/\partial
n\propto K(x)$ is a family-dependent constant. Based on this and previous
findings, we conclude that in CLBLCO not all oxygens contribute holes to the
superconducting and antiferromagnetic order parameters. We also show that
weak-chemical-disorder, as determined from $\Delta \nu _{Q}$, is not playing
a role in the $T_{c}^{max}$ variations. This reinforces previous conclusions
that the changes in $T_{c}^{max}$ between different CLBLCO families is
caused only by variation in the in-plane coupling constant $J$ \cite%
{OferToBe,KanigelPRL02}. Finally, since $\nu _{q}$ is not changing in the
overdoped side as the oxygen level increases, in this side holes are not
added to the planes. The open question in this study is which physical or
chemical property of CLBLCO sets $K(x)$.

This work was funded by the Israeli Science Foundation. We are grateful to
Arkady Knizhnik for his help with the Iodometric titration and to Assa
Auerbach for helpful discussions.

\end{document}